\documentclass[aps,prb,showpacs,amsmath,amssymb,floatfix,twocolumn]{revtex4}
\usepackage{color}
\usepackage{epsfig} \usepackage{bm}
\IfFileExists{srcltx.sty}{\usepackage[active]{srcltx}}

\begin{document}
\bibliographystyle{biblio/prsty}
\title{Time-dependent Transport in arbitrary extended driven tunnel junctions\\}

\author{In\`es Safi}

\affiliation{Laboratoire de Physique des Solides, Universit\'e Paris-Sud, 91405 Orsay, France}

\begin{abstract}
 We develop a very general perturbative theory of time-dependent transport in a weak tunneling junction which is independent of experimental details and on many-body correlated states in the coupled conductors. These can be similar or different, with arbitrary internal or mutual interactions, superconducting correlations, disorder, and coupled to an electromagnetic environment or other quantum systems. The junction can be spatially extended, and is subject, simultaneously, to time-dependent voltage, local magnetic field and modulation of the tunneling amplitudes. All observables at arbitrary frequencies: average current, non-equilibrium admittance  and current correlations can be expressed in a universal way through the out-of-equilibrium DC current only, yielding perturbative time-dependent non-equilibrium fluctuation relations. In particular, charge fluctuations are shown to be universally super-poissonian, and to become poissonian if the junction is driven by a series of Lorentzian pulses. We also generalize, for constant voltage and tunneling, the poissonian shot noise and the fluctuation relation between the derivatives of the noise and the conductance. Thus we provide a compact, general and transparent unifying theory at arbitrary dimension, in contrast with involved derivations based  explicitly on particular models and profiles of a single time-varying field.\end{abstract}


\maketitle
 
{\em Introduction.} A weak tunneling junction between two conductors is the simplest though one of the most useful mesoscopic systems. Its theoretical study is facilitated by perturbative computations with respect to tunneling amplitudes. Among the huge number of its experimental interests, let us recall that: -it is often a building block of hybrid structures -it serves to probe the density of states -it allows to measure the tunneling charge carrier, using the poissonian non-equilibrium shot noise. Weak tunneling, thus high resistance, is also expected to enhance some physical phenomena. This is the case for dynamical Coulomb blockade,\cite{ingold_nazarov}
 when the current is reduced by inelastic tunneling due to exchange of photons with the electromagnetic environment.\cite{ines_saleur_note,note_TLL} Inelasticity can also be induced by coupling to time-dependent (TD) fields, such as the bias voltage $V(t)$, gate voltage or local magnetic field, or classical noisy sources. Then understanding its interplay with inelasticity due to Coulomb interactions and to an electromagnetic environment  is one exciting challenge very little explored so far.\cite{out_green}  In addition, and to the best of our knowledge, no TD profile other than sine or abrupt switch has been treated so far, nor the simultaneous presence of TD barrier modulation and a TD voltage bias $V(t)$. There are also realistic and crucial features which are important to include:  spatial extension of the junction, capacitive coupling, and coupling to other quantum conductors. The aim of the present letter is to develop a very general framework for TD transport including all these ingredients simultaneously, without knowledge of experimental details, nor the underlying Hamiltonian and many-body states. It pursues along the lines Ref.(\onlinecite{ines_eugene}) focussing on the current in periodically driven tunnel junctions.
Two aspects of TD transport can be distinguished and treated here:\cite{buttiker_revue_time}  On one hand, in the DC bias regime, one deals with spontaneous generation through finite frequency non-equilibrium admittance and shot noise at finite frequency.\cite{nazarov_book} On the other hand, TD driving fields\cite{photo_review} generate a TD
%\cite{FCS_TD_tunnel_09,vanevic_nazarov_belzig,glattli_photo} 
 current, whose DC component yields the rectified current or pumped charge. They also affect
current fluctuations, function of two frequencies as well,\cite{lesovik_photo} and of which the (double) DC component, yields the second charge cumulant. Even more, we express for the first time the generalized non-equilibrium admittance depending non-linearily on the TD fields, and whose microscopic exact expression was derived in Ref.(\onlinecite{ines_philippe}). 

We show that all these time-dependent observables can be expressed in a universal manner through the DC tunneling non-equilibrium current. This striking fact is based on a second-order perturbation with respect to a general tunneling operator $\mathcal{T}$, without  recourse to Keldysh technique, but to basic properties of  correlation functions.
In addition to be weak, the unique restriction on $\mathcal{T}$ is that its auto-correlations vanish in absence of tunneling. This amounts to require the absence of pairing correlations of the tunneling excitations, thus, when we specify to a thermal distribution and a superconducting junction, to a negligible supercurrent, ensured by a large capacitance, a magnetic field or dissipation.  As $\mathcal{T}$ nor the global Hamiltonian $\mathcal{H}_0$ without tunneling are specified, our theory is relevant, more generally, to highly resistive strongly correlated conductors, and should be useful in other contexts of perturbation schemes.

Though all these additional extensions are included, specifying some among our universal formulas to either a sine potential at frequency $\Omega$ or a cosine modulation of the barrier allows to recover their form within the Tien-Gordon theory for photo-assisted tunneling,\cite{tucker_rev,buttiker_traversal_time,lesovik_photo} provided the electron charge $e$ is replaced by a general value of the tunneling charge $q$. Derived for independent particles, this theory is based on the side-band transmission scheme: a tunneling electron can exchange $n$ "photons" of  frequency $\Omega$ for all integer $n$ with a probability given by a Bessel function squared. Thus different observables can be obtained by weighing
 their DC values evaluated at  $V$ shifted by $n\Omega/e\hbar$. Indeed, even though the one-particle picture is inappropriate, we show that a similar picture can be extended fully in terms of many-body correlated states. 
 We also obtain formulas never obtained so far, i. e. pioneering even in absence of interactions. 
In addition to works dealing with non-interacting systems,\cite{tucker_rev,buttiker_traversal_time,lesovik_photo} the present theory includes, unifies and goes beyond many other works based on specific models and tedious derivations. \cite{FDT_rogovin_scalapino_74,lee_levitov,FDT_eq_nonlinear_kobayashi_PRB_11,bena,keeling_06_ivanov,sassetti_99_photo,photo_lin_fisher,photo_TLL_two_morais_PRB_06,photo_TLL_extended_morais_PRB_07,photo_gefen_LL,photo_gogolin_TLL_PRL_93,LL_pumping,photo_cheng_TLL_leads_06,photo_cheng_11,photo_komnik_TLL_one_TD_imp,salvay,photo_spin_hall_dolcini,photo_TLL_ring_perfetto_13,kane_fisher_noise,levitov_reznikov,safi_chevallier,photo_crepieux} It applies as well to an arbitrary series of Lorentzian pulses, claimed to generate single-particle excitations free of holes, \cite{keeling_06_ivanov} achieved in a pioneering recent experiment,\cite{glattli_levitons_nature_13} of which we revisit the extension to non-linear conductors. 
%Another exciting field has emerged more recently in connection with generation of single-particle excitations, for instance within the domain of quantum optics with electrons.
%\cite{feve_07,Bocquillon_13_splitter_electrons_demand,Bocquillon_12_HBT_optics} when $V(t)$ is formed by Lorentzian pulses,This choice minimizes the DC charge fluctuations at a given %integer transferred charge fixed by $V$ only. The proposal was claimed to apply to edge states of
 \begin{figure}[htbp]
\begin{center}
\includegraphics[width=6cm]{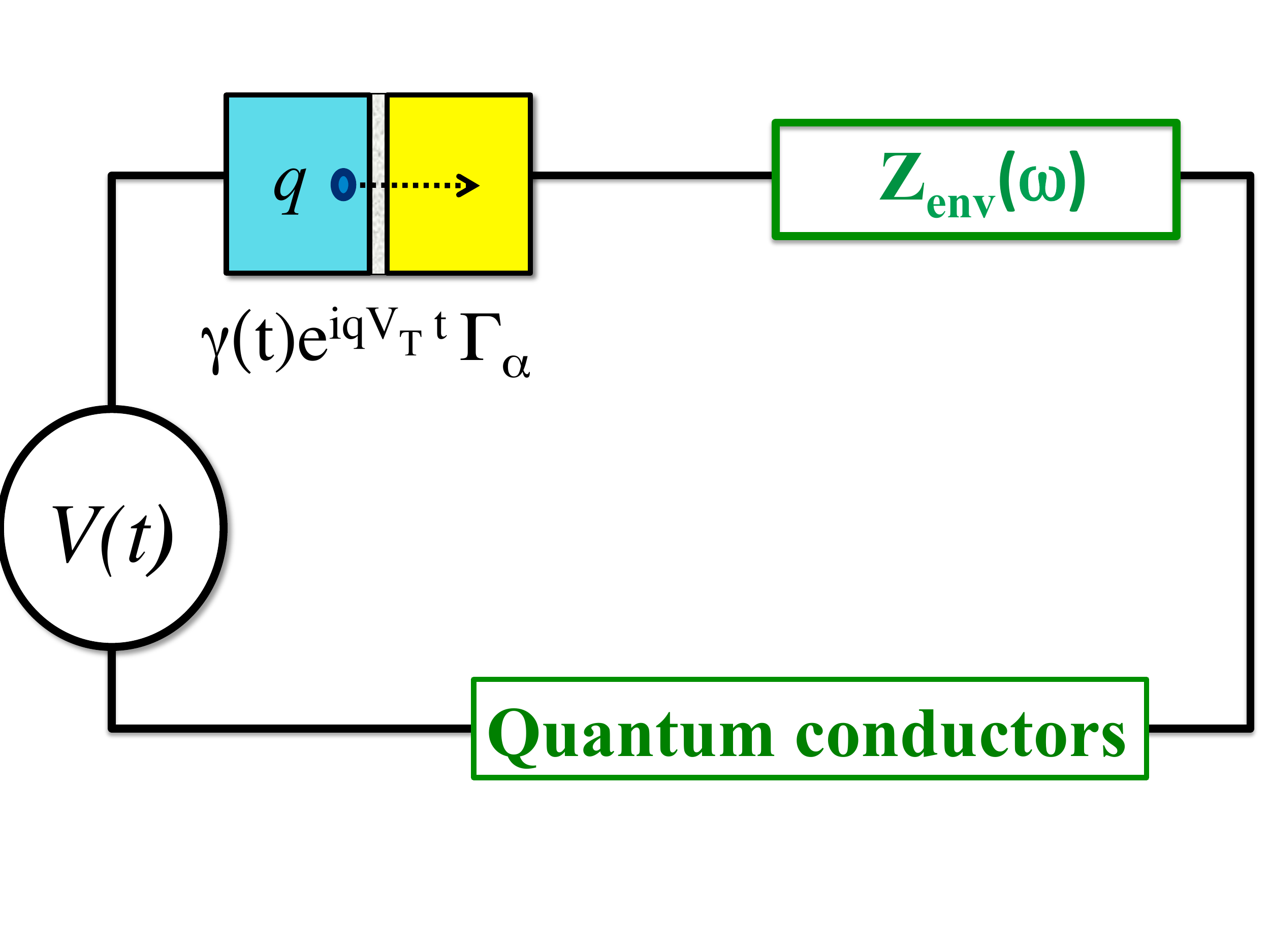}
\caption{\small Extended tunneling between arbitrary similar or different conductors with mutual (thus capacitive) or internal Coulomb interactions, disorder, superconducting correlations, and coupling to other quantum systems and an electromagnetic environment with impedance $Z(\omega)$. Tunneling amplitudes scan depend on many-body states $\alpha$, as well as on time through an arbitrary complex function $\gamma (t)e^{iqV_T t}$, which is the case for TD local magnetic field and gate voltage}\label{setup}
\end{center}
\end{figure}

{\em Model and observables.} Let us consider a tunnel junction between two similar or different interacting/disordered conductors 1 and 2 with mutual interactions, connected to an electromagnetic environment or other quantum systems (see Figure \ref{setup}). We let $k_B=\hbar=1$. The whole quantum circuit is described by a Hamiltonian $\mathcal{H}_0$ in absence of tunneling, required to obey $[Q,\mathcal{H}_{0}]=0$,\cite{ines_eugene} where $Q$ is the charge  operator of the junction. Thus only tunneling, described by an operator $\mathcal{T}$, transfers a charge $q$, {\em i.e.}, $[\mathcal{T},Q]=q\mathcal{T}$,which defines $q$, possibly as an "effective" charge.\cite{q_TLL} 

 We don't have to specify any form for $\mathcal{T}$; it can, for instance, describe tunneling between single or many-body states $\alpha$ with associated TD amplitudes $\Gamma_{\alpha}(t)$, provided the ratio $\Gamma_{\alpha}(t)/\Gamma_{\alpha}(0)=\gamma(t) e^{i qV_T t}$ is state independent. $\gamma(t)$ is an arbitrary complex function, controlled, for instance, by a local gate voltage or a magnetic field; any linear term in its phase $qV_Tt$ is separated, such that stationary regime corresponds to a finite $V_T$ and $\gamma(t)=1$. $V_T$ enters implicitly into a total effective DC voltage $V$, thus $V(t)=V+\tilde{V}(t)$ where $\tilde{V}(t)$ is the ac part. After gauge transformation,\cite{ines_eugene}  the total Hamiltonian reads: $\mathcal{ H}(t)=\mathcal{H}_0+\mathcal{H}_T(t)$, with
\begin{equation}
\mathcal{H}_T(t)= e^{iqVt}\tilde{E}(t)\mathcal{T}+e^{-iqVt}\tilde{E}^*(t)\mathcal{T}^{\dagger}, \label{Hamiltonian} 
\end{equation}
  where $\tilde{E}(t)=\gamma(t) e^{iq\tilde{\phi}(t)} $ and $\partial_t{\tilde{\phi}}(t)=\tilde{V}(t)$. We can show that $\int {d\omega'}\tilde{E}^*(\omega')\tilde{E}(\omega'+\omega) = {2\pi} | \gamma |^2(\omega)$, i. e. the Fourier transform of $|\gamma(t)|^2$. In particular, if $|\gamma(t)|=1$, $|\tilde{E}(\omega')|^2$ can be viewed as the probabililty to exchange photons with an arbitrary frequency $\omega'$, furnished by the effective ac voltage $\tilde{V}(t)+ \partial_t\arg \gamma(t)$.
%\begin{eqnarray}\label{relations}
%-\int \frac{d\omega'}{2\pi} \omega' \tilde{E}^*(\omega')\tilde{E}(\omega'+\omega)&=&q\tilde{V}\otimes R (\omega)+\frac{\omega}2 R(\omega), 
%\end{eqnarray}
%where $\otimes$ refers to integral convolution. 
The whole system is at equilibrium at $t=-\infty$, with a density matrix $\hat{\rho}$, thus the average current reads:
\begin{equation}
\label{current_general} \tilde{I}(qV;t)={\rm Tr}\left[\hat{\rho}\hat I_H(t)\right],
\end{equation}
where the subscript $H$ denotes the Heisenberg representation with respect to $\mathcal{ H}(t)$ of the operator:
$
\hat I(t) =-iq\left[ e^{iqVt}\tilde{E}(t) \mathcal{T}- e^{-iqVt}\tilde{E}^*(t)\mathcal{T}^{\dagger}\right].
$
 The non-equilibrium TD admittance, non-local in time, reads:\begin{eqnarray}\label{admittance_definition}
\tilde{G}(qV;t,t')& =& \frac{\delta \tilde{I}(qV;t)}{\delta \tilde{V}(t')}.
\end{eqnarray}
Non-symmetrized current correlations are defined as:
\begin{equation}\label{Sdefinition}
\tilde{S}(qV;t,t')={\rm Tr}\left[\hat{\rho} \delta\hat I_H(t')\delta\hat I_H(t)\right],\end{equation} 
where $\delta\hat I _H(t)=\hat I_H(t)- \tilde{I}(qV;t)$. $\tilde{G}(qV;t,t')$ was indeed introduced in Ref.(\cite{ines_philippe}), providing a microscopic expression still depending on the driving fields, and is related exactly to $\tilde{S}(qV;t,t')$  (see the Supplemental Material). The double Fourier transforms $\tilde{G}(qV;\omega, \Omega),\tilde{S}(qV;\omega, \Omega)$ are such that
%= \int \int dx ds e^{i \omega x + i \Omega s}  \bar{F} (qV;x, s), $ where $F (qV;t, t') = \bar{F} (qV;x, s)$, with $x = t - t'$ and $s =(t + t')/2$. 
 $\omega + \Omega/2$
and $\omega-\Omega/2$ are conjugate to $t $ and $t'$. Thus $\omega$ becomes the relevant frequency in the stationary regime, while $\Omega$ allows to construct deviations from stationarity. While the dependence of $\tilde{I}$, $\tilde{G}$ and $\tilde{S}$ on $qV$ is explicit, their functional dependence on $q\tilde{V} (t)$ and $\gamma(t)$ is implicit, recalled through the tilde. This is dropped in the stationary regime, {\em i. e.} when $\tilde{V}(t)=0$, $\gamma(t)=1$, for which $\tilde{I}(qV;\Omega)\rightarrow \delta(\Omega)I(qV)$, $\tilde{G}(qV;\omega,\Omega)\rightarrow\delta(\Omega)G(qV;\omega)$ and $S(qV;\omega,\Omega)\rightarrow\delta(\Omega)S(qV;\omega)$. 
%Note that $X_{-}(\omega)$ is real. 
%In the following we will show that, regardless of the form of $\hat{\rho}_0$, $\tilde{I}(qV;t)$ and $\tilde{G}(qV;t,t')$ are fully determined by the out-of-equilibrium DC current characteristics %$I(\omega=qV)$, the unique function keeping track of $\mathcal{H}_0$, $\mathcal{H}_T$ and $\hat{\rho}_0$, all arbitrary. Then, specifying to $\hat{\rho}_0= e^{-\beta \mathcal{H}_0}/Tr(e^{-%\beta \mathcal{H}_0})$  ($\beta=1/T$, $T$ is the temperature), the same striking fact will be shown for $\tilde{S}(qV;t,t')$ (in this case, dependence on $T$ is implicit in all obervables). 

In order to express perturbatively $\tilde{I},\tilde{G},\tilde{S}$ to second order in ${\cal T,T}^\dagger$, we replace $\hat{\rho}$ by $\hat{\rho}_0$ in Eqs.~(\ref{current_general},\ref{Sdefinition}) and apply the standard procedure of expansion of the evolution operator in the interaction representation, a step required only for $\tilde{I}$, as $\tilde{G},\tilde{S}$ are already of second order in  ${\cal T,T}^\dagger$. This leads to the appearance of correlators of $\mathcal{T},\mathcal{T}^{\dagger}$ which are invariant by time translation, as TD is controlled by $\mathcal{H}_0 $ only: $\mathcal{T}(t)=e^{i\mathcal{H}_0 t}\mathcal{T}e^{-i\mathcal{H}_0t}$, and average taken without tunneling, $\langle...\rangle_0={\rm Tr}[\hat{\rho_0} ...]$. We assume that:
\begin{equation}\label{condition}
\langle \mathcal{T}(t)\mathcal{T}(0)\rangle_0=0.
\end{equation}
Thus non-equilibrium dynamics is contained only in $\tilde{E}(t)$, $\tilde{E}^*(t')$, while only two building blocks at equilibrium are needed: $X^>(t)=\langle\mathcal{T}^{\dagger}(t)\mathcal{T}(0)\rangle_0$ and $X^<(t)=\langle\mathcal{T}(0)\mathcal{T}^{\dagger}(t)\rangle_0$. Let us summarize the results through three main facts. First, $X^R(\omega)$, the Fourier transform of $X^R(t)=\theta(t){[X^>(t)- X^<(t)]}$, determines the  DC out-of-equillibrium current in the stationary regime:
\begin{equation}\label{I_DC}
I(qV)=-\frac{ q}{\pi}Re X^R(-qV).
\end{equation}
Secondly, the average current $\tilde{I}(qV;\Omega)$ for the driven junction can be expressed in terms of $X^R$, thus related to $I(\omega=qV)$ using Eq.(\ref{I_DC}) and Kramers Kronig relation. Thirdly, $\tilde{S}(qV;\omega,\Omega)$ can be expressed in terms of $X^>(\omega),X^<(\omega)$. If we specify to $\hat{\rho_0}=e^{-\beta\mathcal{H}_0}/Tr(e^{-\beta \mathcal{H}_0})$, $X^>(\omega)$ and $X^<(\omega)$ become related to $Re X^R(\omega)=-\pi I(-\omega=qV)/q$ by equilibrium FDTs: $X^>(\omega)=-\pi[N(\omega)+1]I(-\omega)/q$, $X^<(\omega)=-\pi N(\omega)I(-\omega)/q$, with $N(\omega)=[e^{\beta\omega}-1]^{-1} $. Therefore, $\tilde{S}(qV;\omega,\Omega)$ can be expressed merely in terms of $I(\omega=qV)$. Now we will give directly these universal expressions, referring to the Supplemental Material for more details.

 {\em Average current and admittance}  To lowest order/ tunneling, Eq.(\ref{current_general}) is  expressed universally as:
 \begin{eqnarray}\label{Iomega_text}
\tilde{I}(qV;\Omega)&=&i\int \int {d\omega'} d\omega"\tilde{E}^*(\omega'-\Omega/2)
\tilde{E}(\omega'+\Omega/2)\nonumber\\&&\frac{(\Omega/2+i\delta)I(qV+\omega"-\omega')}{\omega"^2-(\Omega/2+i\delta)^2},
\end{eqnarray} 
 where the limit of vanishing $\delta$ has to be taken. The rectified current, the easiest to measure experimentally (or transferred charge), obeys a universal relation: \begin{equation}
\label{zerofreqI} \tilde{I}(qV;0)=\int {d\omega'} |\tilde{E}(\omega')|^2 I(qV-\omega').
\end{equation} 
Even though simple and compact, the r. h. s. depends non-trivially on ${V}(t)$ and $\gamma(t)$.\cite{note_linear} This unique formula contains the results of numerous previous works based explicitly on a specific model within much more restrictive framework, for instance the series of works dealing with the Tomonaga-Luttinger model.\cite{sassetti_99_photo,photo_lin_fisher,photo_TLL_two_morais_PRB_06,photo_TLL_extended_morais_PRB_07,photo_gefen_LL,photo_gogolin_TLL_PRL_93,LL_pumping,photo_cheng_TLL_leads_06,photo_cheng_11,photo_komnik_TLL_one_TD_imp,salvay,photo_spin_hall_dolcini,photo_TLL_ring_perfetto_13} Interestingly, one can interpret it by extending the side-band transmission picture to complicated global many-body states. A tunneling charge $q$ has a probability $|\tilde{E}(\omega')|^2$ to absorb (emit) an energy $\omega'<0 (>0)$, inducing a transition between the many-body eigenstates of $\mathcal{H}_0$ spaced by $\omega'$, thus sees an effective voltage $V-\omega'/q$. The total current is given by the superposition of DC currents at $V-\omega'/q$ weighed by $|\tilde{E}(\omega')|^2$. An interesting application\cite{ines_eugene} to a periodic series of pulses of area $\phi_0$, and $\gamma(t)=1$, led to a Josephson type term, $\sin^2(q \phi_0/2)$ in $\tilde{I}(qV;0$. Another one consists into series of Lorentzian pulses; here we focus on a single one, centered around time $t_1$ with width $\tau_1$:
$
qV(t)= {-2\tau_1}/\left({(t-t_1)^2+\tau_1^2}\right),
$
%two situations. First, to the ohmic regime, which can still hold when conductors are strongly interacting, provided the relevant frequencies are in a certain ohmic range,  $I(\omega)%\simeq G\omega$ (quite often, for a thermal global density matrix, this corresponds to $\omega \ll T$). Using Eq.(\ref{relations}), we obtain a compact and general formula from %Eq.~(\ref{Iomega}) : $\tilde{I}(qV;\omega)=G [R(\omega) V+R\otimes\tilde{V} (\omega)]$. If $R(t)=1$, we recover the known result $\tilde{I}(qV;0)=GV$. Secondly, letting
 the DC voltage is given by the surface of the pulse, $V=V_1=2\pi/q$. Then $
 \tilde{E}(qV_1-\omega)=\delta(\omega)-2\tau_1 e^{-\omega (\tau_1+it_1)}\theta(\omega)$
 ($\theta$ is the Heaviside function). In case one has $I(V=0)=0$,\cite{current} Eq.(\ref{zerofreqI}) does not depend on $t_1$, and has a very general expression:
$\tilde{I}(qV_1;0)=4\tau_1^2\int_0^{\infty} {d\omega} e^{-2\omega\tau_1} I(\omega).$
 For instance, for tunneling of fractional charge ($q=\nu e$) between edge states in the FQHE at filling factor $\nu=1/(2n+1)$, and for a thermal distribution such that $2eT\tau_1<<1$,  $\tilde{I}(qV;0)\propto \tau_1^{2(1-\nu)}$.

Now, we focus on the integrated value of the non-equilibrium TD admittance over fast modes, $\tilde{G}(qV;\omega,\Omega=0)=\delta\tilde{I}(qV;\omega)/\delta{V}(\omega)$, which obeys:\begin{eqnarray}\label{Gtilde}
{\rm Re \tilde{G}}(qV;\omega,0)
% &=&\frac q{2\omega}\int\frac{d\omega'}{2\pi}\left[|\tilde{E}(\omega-\omega')|^2-|\tilde{E}(-\omega-\omega')|^2\right] I(qV+\omega')\nonumber\\
&=&\frac{q}{2\omega}\left[\tilde{I}(qV+\omega;0)-\tilde{I}(qV-\omega;0)\right],
\end{eqnarray}
 where the rectified current on the r.h.s are  given by Eq.(\ref{zerofreqI}), thus $\tilde{G}$ depends non-trivially on ${V}(t)$ and $\gamma(t)$ (complex). In the stationary regime, Eq.(\ref{Gtilde}) reduces to that we found previously \cite{ines_eugene,zamoum_12}: ${\rm Re {G}}(qV;\omega)={q}\left[{I}(qV+\omega)-{I}(qV-\omega)\right]/{2\omega}$.

{\bf \it  Current correlations.}  Here we present other main results of the paper: universal TD non-equilibrium FDR for the non-symmetrised current correlations. For that, we need to specify  $\rho_0$ to be thermal (see Supplemental Material). 
Let's start first by the non-equilibrium stationary regime. Then the finite frequency non-symmetrized noise $S(qV;\omega)$ obeys the FDR:
\begin{eqnarray}\label{noise_DC_non}
S(qV;\omega)/\pi q &= &\sum_{\pm}\pm N(\omega\pm qV) I(qV\pm\omega),
\end{eqnarray}
generalizing that based explicitly on the Tomonaga-Luttinger liquid in Ref.(\onlinecite{bena}). It leads to a universal FDR for the symmetrized noise: $\pi  \sum_{\pm}I(qV\pm\omega)\coth \beta (qV\pm\omega)/2$, generalizing its derivation without interactions,\cite{FDT_rogovin_scalapino_74} and with coupling to an electromagnetic environment,\cite{lee_levitov} provided one replaces $e$ by the arbitrary value of the tunneling charge $q$.\cite{q_TLL}
Let us now specify Eq.(\ref{noise_DC_non}) to zero-frequency: $S(qV)=S(qV,0)=\pi q \coth \left({V}/{2T}\right) I(qV).$ This allows us to derive simultaneously, in the most economical and unifying scheme, two important results. First, for $T<<qV$, the poissonian shot noise result:
\begin{equation}\label{DC_noise_DC}
S(qV)=qI(qV).
\end{equation} 
Secondly, the FDR between the derivatives of the noise and the differential conductance at finite temperatures, taking the zero-bias limit after expansion/ $V$: 
\begin{equation}\label{derivatives} 
\frac{dS}{dV}(V=0)=2 T \frac{dG}{dV}(V=0).
\end{equation} 
The quest for such an FDR in nonlinear conductors has been the subject of intensive theoretical and experimental activities recently. Indeed, a lengthy proof of Eq.(\ref{derivatives}) was given in a specific Tunnel junction, without capacitive coupling neither an environment,\cite{FDT_eq_nonlinear_kobayashi_PRB_11} while we have generalized it here straightforwardly, as a direct consequence of Eq.(\ref{noise_DC_non}). Thus  Eq.(\ref{noise_DC_non}) is shown to unify in a transparent and a single shot all these FDRs, which are different simply due to different regimes of temperatures, voltages and frequencies. 

Let us now  turn to a TD driven junction by arbitrary $V(t)$ and complex $\gamma(t)$. The current fluctuations can be expressed fully in terms of the noise generated by a constant voltage, Eq.(\ref{noise_DC_non}), thus in terms of $I$ too:
\begin{eqnarray}\label{FDT2}
\tilde{S}(qV;\omega,\Omega)&=&\int \frac{d\omega'}{2\pi} \tilde{E}(\omega')  \tilde{E}^*(\omega'+\Omega) \nonumber \\&& \times S(qV-\omega';\omega-\Omega/2).
\end{eqnarray}
Symmetrized correlations can be deduced too, as $\sum_{\pm}\tilde{S}(qV;\pm\omega,\Omega)$. 
Let's infer charge fluctuations, obtained by double integration of $\tilde{S}(qV;t,t')$, thus:
\begin{equation}\label{zerofreqS}
\tilde{S}(qV;0,0)= q\int \frac{d\omega'}{2\pi} |\tilde{E}(qV-\omega')|^2\coth\left(\frac{\omega'}{2T}\right)   I(\omega').
\end{equation}
As the two last terms combine into the zero-frequency noise $S(qV=\omega')$, the same interpretation as that of Eq.(\ref{zerofreqI}) holds here. Eq.(\ref{zerofreqS}) reduces to the result derived for noninteracting conductors\cite{lee_levitov,keeling_06_ivanov} if we replace $I(\omega')$ by $\mathcal{T}\omega'$, with $\mathcal{T}$ a weak transmission coefficient, showing that this substitution holds in much more general setups whenever the DC current is linear.
Now let's compare Eq.(\ref{zerofreqS}), specified tp zero temperature, to Eq.(\ref{zerofreqI}). This leads to a universal inequality:
\begin{eqnarray}\label{zerofreqSL}
\tilde{S}(qV;0,0)&\geq&q |\tilde{I}(qV;0)|,
\end{eqnarray}
thus charge fluctuations are super-poissonian (for a thermal density matrix). In particular, they are not bounded by their value in the stationary regime, $S(qV)$: the inequality by Levitov {\it et al}\cite{keeling_06_ivanov} can be recovered only for a linear junction, and for $\gamma(t)=1$, such that the r. h. s. reduces to the DC current, $q\tilde{I}(qV;0)=qI(qV)=GV=S$ (see Eq.(\ref{DC_noise_DC})). Interestingly, for $\gamma(t)=1$, the equality can be reached when $V(t)$ is a series of Lorentzian pulses, as $\tilde{E}(qV-\omega')$ vanishes for either negative or positive $\omega'$, leading to poissonian charge fluctuations:
$
\tilde{S}(qV;0,0)=q |\tilde{I}(qV;0)|.
$

{\it Discussion and conclusion} We have developed a very general framework to deal with TD transport in a tunnel junction independent on experimental details, with a potential validity for other perturbative schemes. Thus many complications, often discarded, can be included here: spatial extension, capacitive coupling, strong correlations, coupling to other quantum conductors or/and an electromagnetic environment. Simultaneous TD bias voltage, gate voltage and local magnetic field are shown to affect in a nontrivial way the finite frequency average current, the non-equilibrium TD admittance, and the asymmetric part of current correlations. Though all are given by compact universal expressions, using only the out-of-equilibrium DC current $I(\omega=qV)$, through which intervene the unspecified global Hamiltonian and density matrix.  If the latter is thermal, we derive universal a TD non-equilibrium FDR for current correlations at arbitrary frequencies. This allowed us to show that charge fluctuations are super-poissonian, and become poissonian when $V(t)$ is a series of Lorentzian pulses and constant tunneling amplitudes.
We have shown that it is possible to extend the side-band transmission picture in terms of many-body correlated eigenstates.This explains why some of the relations here obtained reduce to the same form as those derived within the Tien-Gordon Theory for non-interacting electrons, once we specifiy to a single driving sine field and to $q=e$. Others have not been derived before, even in absence of interactions, such as the finite frequency current and the non-equilibrium TD admittance.

When specified to the stationary regime, universal non-equilibrium FDRs are obtained for the non-symmetrized noise, leading as well to an FDR for the symmetrized noise. Both can be exploited, by comparison to $I(qV)$, to check experimentally whether symmetrized or non-symmetrised noise is measured, as performed by the group of F. Portier. Taking the zero-frequency limit, we have shown the universality of the poissonian DC noise, Eq.(\ref{DC_noise_DC}), as well as the non-equilibrium FDRs between higher derivatives, Eq.(\ref{derivatives}). This gives a unifying framework for all these FDRs which
goes well beyond related works performed independently of each other:\cite{FDT_rogovin_scalapino_74,lee_levitov,kane_fisher_noise,levitov_reznikov,safi_chevallier,FDT_eq_nonlinear_kobayashi_PRB_11} as we don't decouple $\mathcal{H}_0$ neither specify $\mathcal{T}$, we can include capacitive coupling, extended or energy-dependent tunneling, an electromagnetic environment and other quantum conductors.

Let us finally discuss the relevance of this theory, derived at arbitrary dimension, to one-dimensional strongly correlated systems. First, it provides, in a compact, general and transparent way, finite frequency observables, in contrast with much more involved derivations, which moreover, are: -using explicitly the Tomonaga Luttinger Liquid- specifying to a unique driving field -computing only either the rectified current \cite{sassetti_99_photo,photo_lin_fisher,photo_TLL_two_morais_PRB_06,photo_TLL_extended_morais_PRB_07,photo_gefen_LL,photo_gogolin_TLL_PRL_93,LL_pumping,photo_cheng_TLL_leads_06,photo_cheng_11,photo_komnik_TLL_one_TD_imp,salvay,photo_spin_hall_dolcini,photo_TLL_ring_perfetto_13} or current correlations not compared to the DC current.\cite{photo_crepieux} The present theory accomplishes as well other remarkable achievements as: -as it does not require a specific Hamiltonian, it deals with arbitrary series of fractional filling factors in the FQHE and edge reconstruction\cite{note_TLL}-It  is formally valid for both a tunneling or weak backscattering barrier\cite{note_TLL} -It takes systematically into account the realistic extension of tunneling between edge states and their mutual Coulomb interactions, which are unavoidable\cite{extension_TLL} - It shows the universal non-equilibrium FDR in the stationary regime (Eq.(\ref{noise_DC_non})), extending fully Ref.(\onlinecite{bena}) , as well as the DC shot noise, Eq.(\ref{DC_noise_DC})\cite{extension_TLL}-It provides various complementary methods to measure $q$,\cite{q_TLL} in addition to those proposed in Ref.(\onlinecite{ines_eugene}), which will be discussed separately -it provides an appropriate framework to revisit the injection of minimal excitations by Levitov {\it et al}\cite{keeling_06_ivanov}  in the FQHE.

Finally, our theory is promising to address the interplay between inelasticity, non-linearities and decoherence due to Coulomb interactions, and the exchange of photons with TD driving fields and the electromagnetic environment or/and other conductors. It provides powerful general tests which have to be obeyed, more generally, when the highly resistive limit is taken, within one or two impurity problems, quantum dots, or any structure described by an effective energy-dependent transmission. It has many potential applications such as: -pumping-mixing, choosing different periods for the voltages on different sides of the junction, the modulus and phase of $\gamma(t)$ -classical sources of noise, with arbitrary distributions of these driving fields.\\

{\it Acknowledgments}
The author thanks E. Sukhorukov for previous collaboration and useful insight, as well as F. Hekking for discussions and helpful comments on the manuscript. She is grateful to D. Est\`eve, C. D. Glattli, P. Joyez, F. Portier, and B. Reulet for stimulating discussions.

%, the transferred charge in Eq.(\ref{zerofreqI}) depends non-trivially on $\tilde{V}(t)$ (see also Eq.(\ref{zerofreqS})), thus asking how to choose $\tilde{V}(t)$ in order to minimize the noise for a %fixed charge is not a well-posed question.
%\bibliography{biblio/revues,biblio/data_Novembre_13_copie_vendredi_21,biblio/universal}

\begin{thebibliography}{10}
\bibitem{note_TLL}
For one channel conductors either with interactions or coupled to an
  electromagnetic environment, the good transmitting regime can be treated in a
  similar and symmetric manner to the weak tunneling regime, as long as
  transport is considered at low enough frequencies. This is due to the fact
  that high frequency correlations of the backscattering current operator are
  different from those of the total current, as stressed in
  Refs.(\onlinecite{ines_prb_long,dolcini_05,bena,zamoum_12,souquet}), where duality is
  broken. This makes incomplete other related works.
\bibitem{q_TLL}
Apart from simple filling factors $\nu=1/(2n+1)$, for which $q=\nu e$, the
  value of $q$ for different series of $\nu$ is not a well-settled issue. For
  an edge state at simple filling factor $\nu=1/(2n+1)$, $q= e$ or respectively
  $\nu e$. If the edge state is coupled to a Fermi metal, $q=2\nu e/(1+\nu)$
  ($q=e$) in the strong (weak) coupling regime. \cite{ines_schulz_note}In
  quantum wires or carbon nanotubes with arbitrary range and profile of
  interactions, one has, in all regimes, $q=e$, due to the connection to
  reservoirs.\cite{ines_schulz_note} For two superconductors, $\mathcal{T}$
  describes tunneling of Cooper pairs with an effective charge $q=2e$ at
  energies below the gap, and that of quasi-particles with charge $q=e$ at
  energies above the gap.
\bibitem{note_linear}
It is only for a linear DC current and for $\gamma(t)=1$ that the transferred
  charge, as claimed often, is given by the DC component only,
  $\tilde{I}(qV;0)=GV$. Linearity of the DC current can be valid within some
  regimes in an arbitrary highly resistive correlated system, even when the
  scattering approach is not appropriate.
\bibitem{current}
Notice that a vanishing current at zero voltage, $I(0)=0$, is not generally
  true for an arbitrary $\hat{\rho_0}$, but can be shown for the thermal
  distribution.
\bibitem{extension_TLL}
Based explicitly on the Tomonaga-Luttinger model, the paper by Kane and
  Fisher\cite{kane_fisher_noise} deriving the shot noise poissonian result,
  Eq.(\ref{DC_noise_DC}) dealt with a local tunneling, and even claimed no
  general FDR exists at finite frequency, contrary to our findings.
\end{thebibliography}

\bigskip

{\bf \large Supplemental Material}

\bigskip

{\em Exact expressions for the non-equilibrium TD admittance and FDR.} 

Considering an arbitrary conductor described by a TD Hamiltonian and connected to many terminals at TD voltages, we have provided a microscopic formulae for the generalized non-equilibrium TD admittance, depending on the all TD driving fields (it is a matrix indeed).\cite{ines_philippe}  This general formulae was adapted to arbitrary driven Tunnel junction, without requiring weak tunneling, thus
 \begin{eqnarray}\label{admittance_definition_supple}
\tilde{G}(qV;t,t')& =& \frac{\delta \tilde{I}(qV;t)}{\delta \tilde{V}(t')},
\end{eqnarray}
 is given exactly by:
\begin{eqnarray}
  \label{conductance_kubo} \partial_{t'} \tilde{G} (qV;t, t') &= & {i} \theta (t - t') \left\langle \right. [
  \hat{I}_H (t), \hat{I}_H(t')] \left. \right\rangle\nonumber\\
&&- {q}^2\delta (t - t' )
  \left\langle \mathcal{H}_T \left( t \right)
  \right\rangle.
\end{eqnarray}
The consequence of Eq.(\ref{conductance_kubo}) is an exact FDR for the asymmetric part of the non-symmetrized current correlations, Eq.(\ref{Sdefinition}), 
 which verify:
\begin{eqnarray}
S^-(qV;t,t')=\tilde{S}(qV;t,t')-\tilde{S}(qV;t',t)=\nonumber \\
\hbar\left[\partial_{t'}G(qV;t,t')-\partial_t G(qV;t,t')\right].
\label{FDT1}\end{eqnarray}
Once Fourier transformed, this yields
\begin{eqnarray}
\tilde{S}(qV;\omega, \Omega) - \tilde{S}(qV;-
\omega, \Omega)  =   \nonumber \\ \left( \Omega/2 - \omega \right)
\tilde{G} (qV;\omega, \Omega)  - \left(\Omega/2 +\omega\right)
\tilde{G}(qV;-\omega, \Omega). \label{compact}
\end{eqnarray}
Here the double Fourier transform of $F(qV;t,t')$ is defined as:
$F(qV;\omega,\Omega)= \int \int dx ds e^{i \omega x + i \Omega s}  \bar{F} (qV;x, s), $ where $F (qV;t, t') = \bar{F} (qV;x, s)$, with $x = t - t'$ and $s =(t + t')/2$. 

Nevertheless, it is only in the stationary regime that this theory has been used so far, leading to a non-equilibrium admittance which depends on the DC voltages and a single frequency, computed for the first time in a three-terminal geometry within the Tomonaga-Luttinger Liquid model with leads\cite{bena} or exactly (without leads) at the specific interaction parameter $1/2$.\cite{zamoum_12} Other works have followed as well, mainly concerning the Kondo problem in a two-terminal geometry.\cite{admittance_FF_dot_wolfle_PRB_13,admittance_FF_kondo_andergassen_PRB_13}
The advantage of the present theory, restricted to weak tunneling regime, is the opportunity to compute the generalized TD admittance and current fluctuations for the driven junction.

{\em Perturbative expressions for the average current,  non-equilibrium TD admittance and current fluctuations.}

Let's first compute the average current to second order with respect to $\mathcal{T}$.  \begin{eqnarray}\label{Iomega}
\tilde{I}(qV;\Omega)&=&-{q}\int \frac{d\omega'}{2\pi} \tilde{E}^*(\omega')
\tilde{E}(\omega'+\Omega)\nonumber\\&&\left[X^R(\omega'+\Omega-qV)+X^{R*}(\omega'-qV)\right].
\end{eqnarray} 
We recall that $X^R(t)=\theta(t)\langle[\mathcal{T}^{\dagger}(t),\mathcal{T}(0)]\rangle_0$. In the stationary regime, $\tilde{I}(qV;\Omega)\rightarrow \delta(\Omega)I(qV)$ with:
\begin{equation}\label{I_DC_supp}
I(qV)=-\frac{ q}{\pi}Re X^R(-qV).
\end{equation}
If we specify first to a vanishing frequency, thus express the rectified current, one has directly a compact equation (given in the main text):
\begin{equation}
\label{zerofreqI_suppl} \tilde{I}(qV;0)=\int {d\omega'} |\tilde{E}(qV-\omega')|^2 I(\omega').
\end{equation} 
For a finite frequency, we need to use Eq.(\ref{I_DC_supp}) together with the  Kramers-Kronig relation such that we express fully $X^R$ in terms of $I$:
\begin{eqnarray}\label{kramers}
X^{R}(\omega)&=& -\frac{ i\pi}{q}\int d\omega' \frac {I(\omega')}{\omega'+\omega+i\delta}.\nonumber\\&&= -\frac{ \pi}{q}\left[I(-\omega)+i P\int d\omega' \frac {I(\omega')}{\omega'+\omega}\right],
\end{eqnarray}
where the limit of vanishing $\delta$ has to be undertaken in the first equation, which is more convenient to adopt here. Substitution into Eq.~(\ref{Iomega})  and additional algebraic steps yield the universal expression for $\tilde{I}(qV,\Omega) $ (given in the text):
\begin{eqnarray}\label{Iomega_suppl}
\tilde{I}(qV;\Omega)&=&i\int \int {d\omega'} d\omega"\tilde{E}^*(\omega'-\Omega/2)
\tilde{E}(\omega'+\Omega/2)\nonumber\\&&\frac{(\Omega/2+i\delta)I(qV+\omega"-\omega')}{\omega"^2-(\Omega/2+i\delta)^2},
\end{eqnarray} 
thus  is determined fully and universally by $I(\omega=qV)$ and $\tilde{E}(t)$. 
Now we can derive the generalized non-equilibrium TD admittance in two ways. Either we use the exact formuae in Eq.(\ref{conductance_kubo}) with the average of the commuter taken now without tunneling, i. e. related to $X^R$:
\begin{eqnarray}
\partial_{t'}\tilde{G}(qV;t,t')=-{2q^2}\Im\mbox{m } \int dt"e^{-iqV(t-t")} \times \nonumber \\ \tilde{E}^*(t)\tilde{E}(t")X^R(t-t")\left[\delta(t-t')-\delta(t"-t')\right].\label{partialG}
\end{eqnarray}
Alternatively, we can use the perturbative computation of the TD current, Eq.~(\ref{Iomega}), and take its  differential with respect to $\tilde{V}(t)$. For that, we can use a general exact result obeyed by the Fourier transform of $\tilde{E}(t)$ :
\begin{eqnarray}\label{properties}
\frac {\delta \tilde{E}(\omega)}{\delta \tilde{V}(\omega')}&=&-\frac{q}{\omega'}\tilde{E}(\omega-\omega').
\end{eqnarray}
By recovering the same expression as the double Fourier transform of Eq.(\ref{partialG}), we derive a non-trivial verification of the general TD non-linear transport theory in Ref.(\onlinecite{ines_philippe}).
Let's now consider current fluctuations,which have the perturbative expression:
\begin{eqnarray}\label{Somegaomega'}
\tilde{S}(qV;\omega,\Omega)=q^2 \int d\omega'\tilde{E}(\omega')  \tilde{E}^*(\omega'+\Omega) \nonumber\\ \left[X^<(-qV+\omega+\omega'-\Omega/2)+ X^>(-qV - \omega +\omega'+\Omega/2)\right],
\end{eqnarray}
which obeys the exact FDR given by Eq.(\ref{compact}), with $\tilde{G}$ given by Eq.(\ref{partialG}).\cite{ines_philippe} Thus the asymmetric part is expressed as well through the DC current. Now we proceed further by specifying $\hat{\rho_0}=e^{-\beta\mathcal{H}_0}/Tr(e^{-\beta \mathcal{H}_0})$. This allows us to benefit from the detailed balance relation:
$X^>(\omega)=e^{-\beta\omega}X^<(\omega)$, and express both $X^{>,<}$ through equilibrium FDTs:
$X^>(\omega)=[N(\omega)+1]X_-(\omega)$, $X^<(\omega)=N(\omega)X_-(\omega)$, where $N(\omega)=[e^{\beta\omega}-1]^{-1} $.  This allows to recover the general expression for $\tilde{S}(qV;\omega,\Omega)$:
\begin{eqnarray}\label{FDT2_supp}
\tilde{S}(qV;\omega,\Omega)&=&\int \frac{d\omega'}{2\pi} \tilde{E}(\omega')  \tilde{E}^*(\omega'+\Omega) \nonumber \\&& \times S(qV-\omega';\omega-\Omega/2),
\end{eqnarray}
while the symmetrized part can be deduced through by $\sum_{\pm}\tilde{S}(qV;\pm\omega,\Omega)$. Here $S(qV;\omega)$ is shown to be determined by $I$,
\begin{eqnarray}\label{noise_DC_non_supp}
S(qV;\omega)/\pi q &= &\sum_{\pm}\pm N(\omega\pm qV) I(qV\pm\omega),
\end{eqnarray}
One can check that this finite-frequency noise under a DC voltage is obtained from Eq.(\ref{FDT2_supp}) in the stationary regime, as one takes the limit $\tilde{E}(\omega') =\delta(\omega')$.

\end{document}